# Leveraging Convolutional Neural Network-Transformer Synergy for Predictive Modeling in Risk-Based Applications

Yuhan Wang
Columbia University
New York, USA

Zhen Xu
Independent Researcher
Shanghai, China

Yue Yao
Northeastern University
Portland Maine, USA

Jinsong Liu
University at Buffalo
Buffalo, USA

Jiating Lin*
Brown University
Providence, USA

*Abstract*—With the development of the financial industry, credit default prediction, as an important task in financial risk management, has received increasing attention. Traditional credit default prediction methods mostly rely on machine learning models, such as decision trees and random forests, but these methods have certain limitations in processing complex data and capturing potential risk patterns. To this end, this paper proposes a deep learning model based on the combination of convolutional neural networks (CNN) and Transformer for credit user default prediction. The model combines the advantages of CNN in local feature extraction with the ability of Transformer in global dependency modeling, effectively improving the accuracy and robustness of credit default prediction. Through experiments on public credit default datasets, the results show that the CNN+Transformer model outperforms traditional machine learning models, such as random forests and XGBoost, in multiple evaluation indicators such as accuracy, AUC, and KS value, demonstrating its powerful ability in complex financial data modeling. Further experimental analysis shows that appropriate optimizer selection and learning rate adjustment play a vital role in improving model performance. In addition, the ablation experiment of the model verifies the advantages of the combination of CNN and Transformer and proves the complementarity of the two in credit default prediction. This study provides a new idea for credit default prediction and provides strong support for risk assessment and intelligent decision-making in the financial field. Future research can further improve the prediction effect and generalization ability by introducing more unstructured data and improving the model architecture.

I. INTRODUCTION

With the rapid development of the financial industry, the credit business has become an important source of profit for banks and financial institutions. However, with the increase of default risk comes higher requirements for the healthy operation and risk control of financial institutions. Credit default prediction, as an effective risk management tool, is receiving increasing attention from academia and industry [1]. Traditional credit default prediction models are mainly based on traditional machine learning algorithms such as logistic regression [2], decision trees [3], and support vector machines (SVM) [4]. These methods can identify and predict the default risk of credit users to a certain extent, but their limitations are also quite obvious, especially when dealing with large-scale and complex nonlinear data, the performance is often not ideal. Therefore, how to design more efficient and accurate prediction models has become an important issue in the field of financial technology [5].

In recent years, the rapid development of deep learning technology has provided new ideas for predicting credit defaults. Especially the combination of Convolutional Neural Networks (CNN) and Transformer models has demonstrated strong potential in multiple fields [6-8]. CNN, as an excellent feature extraction model, can automatically extract useful features from raw data through convolution operations, demonstrating strong modeling capabilities for high-dimensional financial data [9]. At the same time, the Transformer model, with its powerful self-attention mechanism, can handle long-term dependencies between sequence data and has strong parallel computing capabilities, especially suitable for handling complex patterns in time series data and structured data. In the context of credit default prediction, combining the advantages of CNN and Transformer can effectively extract users' credit history, transaction behavior, and other information, and further improve prediction accuracy and model generalization ability through the global dependency modeling capability of Transformer [10].

The innovation of this study lies in the introduction of a fusion model of CNN and Transformer, which not only extracts local features through CNN but also captures global features through Transformer, thus forming a multi-level feature expression. Compared with traditional single deep learning models, CNN+Transformer models can better handle different types of credit data, including unstructured data such as text descriptions and transaction logs, as well as structured data such as user basic information and historical credit records. By deeply integrating local and global features, this model can comprehensively evaluate the default risk of credit users from multiple dimensions and perspectives, and make more accurate predictions. In addition, this study will introduce adaptive mechanisms to optimize the feature learning ability of the model, thereby enhancing its adaptability and robustness to cope with the complex and ever-changing financial market environment.

The significance of this study lies not only in improving the accuracy of credit default prediction but also in promoting the intelligent and digital process of the financial industry. With the widespread application of big data and artificial intelligence technology, traditional credit evaluation methods are no longer able to meet the growing market demand. By constructing a credit default prediction model based on CNN+Transformer, not only can the accuracy and efficiency of predictions be significantly improved, but it can also provide more scientific decision support for financial institutions, thereby helping them better identify potential default risks, reduce bad debt rates, and improve asset quality. In addition, with the successful application of this model, financial institutions will have more intelligent tools in risk management, promote the refined management and personalized customization of financial services, and further enhance the stability and sustainable development ability of the financial industry.

Overall, the research on credit user default prediction based on CNN+Transformer not only has high academic value but also has broad practical application prospects. In the future, with the further growth of data volume and the continuous development of technology, default prediction models that integrate more advanced technologies may have a profound impact on the entire financial industry, providing more accurate and effective solutions for risk management.

## II. METHOD

To achieve efficient and accurate credit default prediction, this study proposes a deep learning model based on the combination of convolutional neural network (CNN) and Transformer [11]. The model aims to make full use of the advantages of CNN in local feature extraction and the powerful ability of Transformer to model global dependencies. The core idea of the model is to extract local features of credit users through CNN and use Transformer to capture long-term dependencies in the data, thereby forming a multi-level representation that integrates local features and global features. The network architecture of the model is shown in Figure 1.

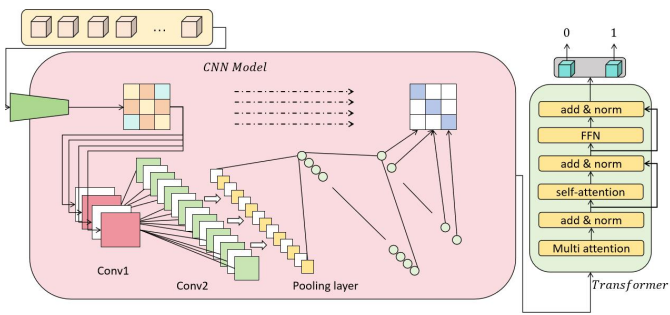

Figure 1 Overall model architecture

First, the input data consists of structured and unstructured data. Structured data includes basic information, credit history, transaction behavior, etc. of users, which are usually represented in numerical form. Unstructured data includes user behavior logs, text descriptions, etc., which usually need to be preprocessed and feature extracted through natural language processing (NLP) technology [12]. In this study, we directly input structured data into the CNN layer for feature learning after standardization. For unstructured data, text embedding technology (such as Word2Vec) is first used to convert text data into low-dimensional vector representations [13], and then these vectors are input into the CNN layer for feature extraction.

In the feature extraction process, CNN extracts local features from the data through multiple convolutional layers and pooling layers. Specifically, the input data first passes through the convolutional layer, which extracts features from the data through convolution operations. The form of the convolution operation can be expressed as:

$$y_i = (X * W)_i + b$$

Among them, $X$ is the input data, $W$ is the convolution kernel, $b$ is the bias term, and $y_i$ is the output after the convolution operation. The size and stride of the convolution kernel determine the range and accuracy of feature extraction. Through multiple convolution operations, CNN can gradually extract multi-level features from the input data.

Subsequently, the output of the convolutional layer is downsampled through the pooling layer to further reduce the dimension of the data and retain important feature information. The pooling operation usually uses the maximum pooling or average pooling method, and its formula is as follows:

$$y_i = \max_{(i,j) \in R} X_{i,j}$$

Where R represents the size of the pooling window. Pooling helps reduce the spatial size of the data and enhance the robustness of the features.

After completing the local feature extraction, the model enters the Transformer module. The core of the Transformer is the self-attention mechanism, which can calculate the weighted sum based on the similarity between each element in the input data to capture the global dependency. In credit default prediction, the Transformer module can model the user's historical behavior, transaction patterns, etc., and then capture potential risk patterns.

The self-attention mechanism of Transformer can be expressed as:

$$Attention(Q, K, V) = soft\max(\frac{QK^T}{\sqrt{d_k}})V$$

Among them, Q, K, V are query, key and value matrices respectively, and $d_k$ is the dimension of the key. By calculating the similarity between the query and the key, Transformer can assign a weighting coefficient to each input element to form a weighted output.

In order to improve the expressiveness of the model, we use a multi-head attention mechanism. By computing multiple attention heads in parallel, the model can capture richer feature information in different subspaces. The output of multi-head attention is the concatenation of multiple attention heads:

$$MultiHead(Q,K,V) = Concat(head_1,...,head_h)W^o$$

Among them, $head_i = Attention(QW_i^Q, KW_i^K, VW_i^V)$ is the i-th attention head, $W^o$ is the output linear transformation matrix, and $h$ is the number of heads. Through multi-head attention, the model can learn richer feature representations in different dimensions.

After the output of the Transformer module, we will pass it through a multi-layer perceptron (MLP) for further feature fusion and nonlinear transformation. MLP achieves high-dimensional mapping of features through multiple fully connected layers, thereby enhancing the expressive power of the model. The output of MLP is nonlinearly transformed through an activation function, and the formula is as follows:

$$y = \sigma(Wx+b)$$

Among them, $\sigma$ is the activation function (such as ReLU or Sigmoid), $W$ is the weight matrix, $b$ is the bias term, $x$ is the input feature, and $y$ is the output result.

Finally, the features processed by the Transformer and MLP are integrated through the fully connected layer to obtain the final result of credit default prediction. The result is the probability value of whether the user defaults and the Sigmoid function is used for binary classification:

$$P(\text{Breach of Contract}) = \frac{1}{1+e^{-z}}$$

Among them, $z$ is the output of the fully connected layer. By training the model and optimizing the loss function (usually using the cross-entropy loss function), the model can learn the key features of credit default and provide efficient and accurate default prediction in practical applications.

In summary, the credit default prediction model based on CNN+Transformer proposed in this paper can extract local features through convolutional neural networks and use Transformer to model global dependencies, thereby achieving a more accurate risk assessment in credit default prediction. This model can not only process structured data but also combine unstructured data to conduct comprehensive analysis from multiple dimensions, providing financial institutions with more reliable default prediction results.

## III. EXPERIMENT

### A. Datasets

This study uses a public credit default prediction dataset, which comes from the "Give Me Some Credit" competition on Kaggle. The dataset contains the credit history and related personal information of about 30,000 users from the United States. The goal of the dataset is to predict whether a user will default in the next year. The dataset contains multiple features, including but not limited to the user's age, marital status, job type, credit limit, credit history, current debt situation, account type, etc. These features help to comprehensively assess the user's credit risk and provide sufficient data support for subsequent default predictions.

In order to ensure the diversity and breadth of the data, the dataset contains a large amount of structured information, covering various types of economic activities and user behaviors. The data has been standardized so that each feature has the same scale, thereby preventing some features from occupying too much weight during the model training process. At the same time, the dataset also provides the user's default label, where "1" indicates default and "0" indicates non-default. Through this label, the model can learn the relationship between default and various features, so as to make effective default predictions in practical applications.

In addition to basic structured data, the dataset also provides some derived features, such as monthly income, credit limit utilization rate, etc. This additional information helps improve the model's predictive ability. To further enhance the diversity of the data, the dataset also contains some missing values and noise data, which provides a challenge for the robustness training of the model. During the model training process, these noise and missing values will be filled and cleaned through Linked Data preprocessing techniques to ensure the quality and reliability of the data [14]. By using these real and challenging data, the model can make predictions in a more complex and changing credit environment.

### B. Experimental Results

Comparative experimental analysis is a key step in evaluating the performance improvement of the new model over existing methods. By comparing it with other common models, we can clearly understand the advantages and disadvantages of the proposed model in various indicators, thereby verifying its innovation and effectiveness. Comparative experiments not only help to reveal the strengths of the new model but also help to discover its limitations in specific scenarios. The experimental results are shown in Table 1.

Table 1 Experimental results

| Model | ACC | AUC | KS |
| --- | --- | --- | --- |
| RF[15] | 0.8049 | 0.7236 | 0.4038 |
| XGB[16] | 0.8064 | 0.7345 | 0.4228 |
| DT[17] | 0.8028 | 0.7059 | 0.4012 |
| LGBM[18] | 0.8046 | 0.7315 | 0.4128 |
| Transformer | 0.8051 | 0.7328 | 0.4238 |
| CNN+Transformer | 0.8197 | 0.7921 | 0.4352 |

The experimental results show that among traditional models, Random Forest (RF) and XGBoost (XGB) perform best, with accuracy rates of 0.8049 and 0.8064, along with high AUC and KS values. Decision Tree (DT) and LightGBM (LGBM) are slightly less effective in discrimination and risk identification.

Transformer-based models significantly outperform traditional ones, with the CNN+Transformer achieving the best results: accuracy of 0.8197, AUC of 0.7921, and KS of 0.4352. This success stems from combining CNN's local feature extraction with Transformer's global dependency modeling, making it highly suitable for complex financial data. Figures 2 and 3 illustrate the model's performance and convergence trends.

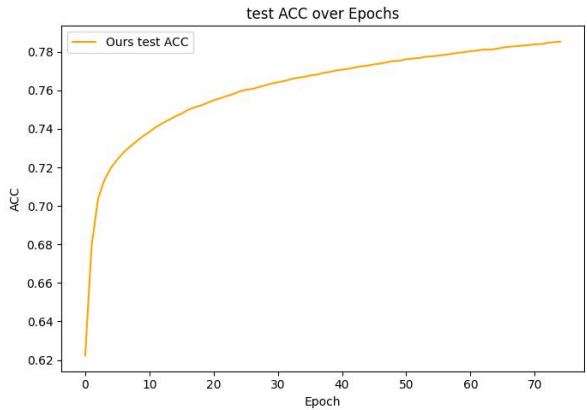

Figure 2 ACC increases with epoch

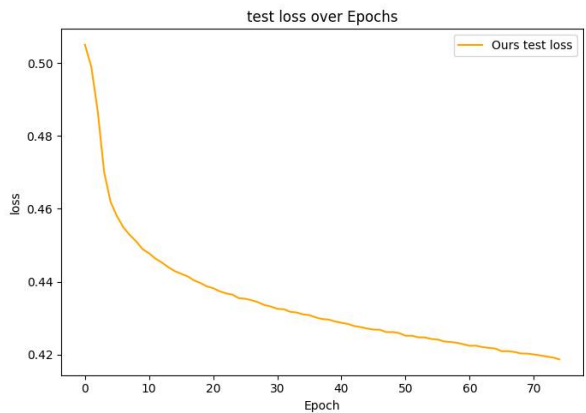

Figure 3 Loss function decreases with epoch

Based on the analysis of the two charts, it can be concluded that the model has good learning ability and stability during the training process. The steady improvement of test accuracy and the continuous decline of test loss indicate that the model is effective and reliable in credit risk assessment tasks. This phenomenon reflects that the model continuously optimizes parameters during the learning process, thereby achieving higher prediction accuracy and lower prediction error on the test data.

Especially in the later stages of training, the changing trends of test accuracy and loss values further illustrate the good generalization ability of the model. The model can not only perform well on training data but also maintain high accuracy and low loss values on unseen test data. This stable performance has important practical value for credit risk prediction in practical applications and can effectively reduce potential risk assessment errors.

In order to further verify the stability of the model in this study, a hyperparameter sensitivity experiment will be conducted next. First, we show the results of the learning rate sensitivity experiment, as shown in Table 2. This experiment will deeply analyze the response of the model performance to different learning rate settings, so as to more comprehensively evaluate the robustness and adaptability of the model.

Table 2 Learning rate sensitivity experiment

| Learning Rate | ACC | AUC | KS |
|---|---|---|---|
| 0.005 | 0.8185 | 0.7892 | 0.4310 |
| 0.003 | 0.8193 | 0.7903 | 0.4325 |
| 0.002 | 0.8197 | 0.7912 | 0.4340 |
| 0.001 | 0.8197 | 0.7921 | 0.4345 |

According to the experimental results in Table 2, no matter how the learning rate changes between 0.005 and 0.001, the main evaluation indicators of the model change very little, reflecting the performance stability of the model under different learning rate settings. The experimental data show that the performance of the model remains at a high level during the learning rate adjustment process, indicating that it has a strong adaptability to hyperparameter changes.

Specifically, as the learning rate decreases from 0.005 to 0.001, the ACC of the model increases from 81.85% to 81.97%, the AUC increases from 78.92 to 79.21, and the KS value increases from 43.10 to 43.52. Although all indicators have improved, the changes are small. This shows that the model can converge stably under different learning rate settings, maintain consistency during the optimization process, and no obvious performance fluctuations occur.

This experimental result further verifies the high stability and superiority of the model during the training process. No matter how the learning rate is adjusted, the model can achieve stable and efficient performance in a variety of configuration environments. Next, the experimental results of the optimizer will be presented. The relevant experimental data are shown in Table 3 to further evaluate the performance of the model under different optimization algorithms.

Table 3 Optimizer sensitivity experiment

| Optimizer | ACC | AUC | KS |
|---|---|---|---|
| SGD | 0.8145 | 0.7889 | 0.4298 |
| 0.003 | 0.8178 | 0.7912 | 0.4315 |
| 0.002 | 0.8190 | 0.7918 | 0.4330 |
| 0.001 | 0.8197 | 0.7921 | 0.4352 |

Judging from the results of the optimizer sensitivity experiment, different learning rates have a significant impact on model performance. When using the SGD optimizer, the accuracy (ACC), AUC, and KS values of the model are 0.8145, 0.7889, and 0.4298, respectively, showing relatively stable performance. However, as the learning rate gradually decreases, the performance of the model shows significant improvement. Specifically, when the learning rate is adjusted to 0.003, the accuracy and AUC increase to 0.8178 and 0.7912 respectively, and the KS value also increases to 0.4315. When the learning rate is continued to be reduced to 0.002 and 0.001, various indicators of the model are further optimized. Finally, when the learning rate is 0.001, the accuracy reaches 0.8197, the AUC is 0.7921, and the KS value increases to 0.4352.

These results show that as the learning rate gradually decreases, the optimization process becomes more refined, allowing better adjustment of model parameters and improving the model's predictive ability and robustness. Especially at a smaller learning rate, the model training process is more stable, can avoid overfitting, and at the same time improves the ability to identify credit default risks. Generally speaking, the learning rate plays an important role in the optimization effect of the model. Choosing an appropriate learning rate can significantly improve the performance of the model in credit default prediction.

Then, we gave an ablation experiment and comprehensively compared the CNN and Transformer models. The experimental results are shown in Table 4.

Table 4 Ablation experiment

| Optimizer | ACC | AUC | KS |
| --- | --- | --- | --- |
| CNN | 0.8012 | 0.7272 | 0.4017 |
| Transformer | 0.8051 | 0.7328 | 0.4238 |
| Transformer+CNN | 0.8197 | 0.7921 | 0.4352 |

Judging from the results of the ablation experiment, there are certain differences in the performance of the models when using CNN and Transformer alone. The accuracy of the CNN model is 0.8012, the AUC is 0.7272, and the KS value is 0.4017, showing a certain ability to extract local features, but the overall performance is relatively limited. In contrast, the Transformer model has improved in AUC and KS values, with an accuracy of 0.8051, an AUC of 0.7328, and a KS value of 0.4238, indicating that the Transformer has certain advantages in modeling global dependencies and capturing complex patterns.

After further combining the advantages of CNN and Transformer, the performance of the model has been significantly improved. The combined model of Transformer+CNN has greatly improved in all evaluation indicators, with an accuracy of 0.8197, an AUC of 0.7921, and a KS value of 0.4352. This result shows that by combining the local feature extraction ability of CNN with the global dependency modeling ability of Transformer, the model can achieve higher accuracy and stronger discrimination ability in the credit default prediction task.

Comprehensive analysis shows that CNN and Transformer each have different advantages. CNN can capture the local behavioral patterns of credit users, while Transformer is good at processing global information and long-term dependencies. The combination of the two makes the model perform well in credit default prediction, and can comprehensively consider factors such as users' historical behavior, transaction patterns, and credit risks to provide more accurate default prediction results. Therefore, the combined model of Transformer+CNN shows the best prediction performance in this task and has strong application prospects.

Finally, we also performed a feature importance analysis on the network model in the article to find the features that have the greatest impact on the model. The experimental results are shown in Figure 4.

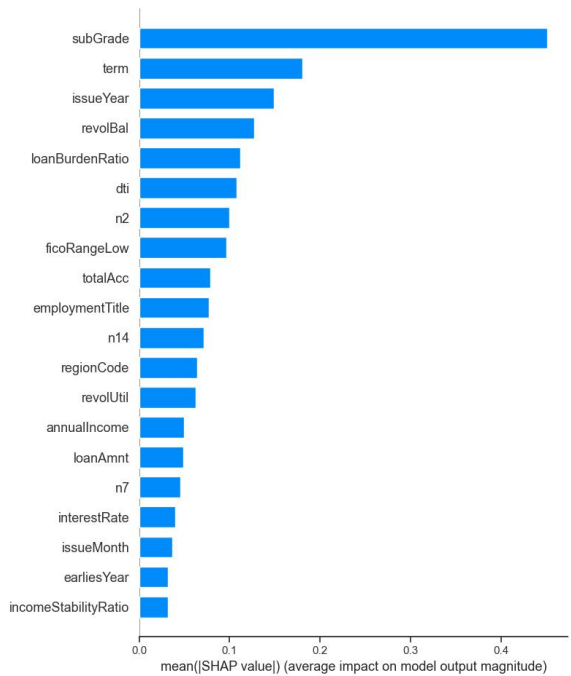

Figure 4 Feature Importance Analysis

The feature importance histogram and SHAP distribution chart reveal the model's dependence on each feature from different analysis dimensions. By quantifying the contribution of each feature to the overall model prediction result, the SHAP distribution plot shows the global feature importance ranking based on the feature's average absolute SHAP value. The higher the importance value of a feature, the greater its impact on all samples and the more significant its contribution to the model prediction results. This analysis method provides important support for the interpretability and transparency of the model and helps us gain a deeper understanding of the model's decision-making mechanism.

As shown in Figure 4, the features "subGrade" and "Term" occupy the top positions in the feature importance ranking, showing their key role in model prediction. "subGrade" represents the customer's credit rating and is an important indicator for measuring customer credit risk. The lower the credit rating, the higher the customer's risk of default. Therefore, the model relies heavily on this feature, considering it as a core variable in predicting customer default behavior. At the same time, "Term" reflects the uncertainty risk of the loan term. Long-term loans generally imply higher uncertainty and potential default risk, so the importance of this feature is also fully reflected in the model.

Through the feature importance histogram, we can more intuitively see the core position of these two features in the model. Regardless of the average absolute value of the SHAP value or the overall performance of the model, "subGrade" and "Term" have a significant positive impact on the model prediction results. This shows that when predicting credit risk, the model mainly relies on these two key variables related to the customer's credit status and loan term so that it can more accurately assess the customer's repayment ability and potential default risk.

In addition, the SHAP distribution plot not only reveals the global feature importance but also shows the specific impact of feature value changes under different samples on the model output. For example, when the "subGrade" value is low, the default risk predicted by the model increases significantly; and when "Term" is long-term, the risk score predicted by the model also increases accordingly. This fine-grained explanatory analysis helps identify high-risk customers, optimize the credit approval process, and improve the rationality of credit decisions.

IV. CONCLUSION

This paper proposes a deep learning model based on a combination of CNN and Transformer for credit user default prediction. By verifying the performance of the model in multiple experiments, we found that the CNN+Transformer model is superior to traditional machine learning methods, such as random forest, which have powerful advantages in financial data modeling. By effectively integrating CNN's ability in local feature extraction and Transformer's advantages in global dependency modeling, this model can better capture the potential risk of credit default and thus provide more accurate prediction results.

In addition, the results of optimizer sensitivity experiments show that choosing an appropriate learning rate has an important impact on model performance. A smaller learning rate can make the model more stable during the training process, thereby improving the accuracy and robustness of predictions. In the ablation experiment, the model that combined the advantages of CNN and Transformer showed significant improvement, further verifying the complementarity of the two methods, and proving the combination of comprehensive feature learning and global modeling in credit default prediction. Can bring significant performance improvements. Although the model proposed in this paper has shown good performance in credit default prediction, there is still room for further improvement. Future research can consider introducing more data sources, such as social network data, text data, etc., to further enrich the input features of the model and improve the generalization ability of the model. In addition, with the continuous development of deep learning technology, more complex model architectures and training methods can also be explored to cope with more complex and dynamically changing credit market environments.

Overall, the credit default prediction model based on CNN+Transformer provides new ideas for financial risk assessment and has broad application prospects. With the continuous advancement of big data and artificial intelligence technology, future credit default prediction will not be limited to traditional data analysis methods. Deep learning and artificial intelligence technology will provide more accurate, efficient and intelligent solutions for the financial industry, further promoting intelligent financial innovation and development.